\documentclass[hyper]{JHEP} 

\usepackage{epsfig}

\newcommand\fverb{\setbox\pippobox=\hbox\bgroup\verb}

\newcommand\fverbdo{\egroup\medskip\noindent%

            \fbox{\unhbox\pippobox}\ }

\newcommand\fverbit{\egroup\item[\fbox{\unhbox\pippobox}]}

\newbox\pippobox

\title{Remark About Scaling Limit of
ABJ Theory}
\author{J. Kluso\v{n}\\
Department of
Theoretical Physics and Astrophysics\\
Faculty of Science, Masaryk University\\
Kotl\'{a}\v{r}sk\'{a} 2, 611 37, Brno\\
Czech Republic\\

E-mail: \email{klu@physics.muni.cz}}
\preprint{  \hepth{0902.4122}}

 \abstract{We generalize the suggestion
 presented in
arXiv:0806.3498 that the $3d \quad N=8$
superconformal $SU(N)$
Chern-Simons-matter theory of
Lorentzian Bagger-Lambert-Gustavson
type (L-BLG) can be obtained through
the scaling limit from $N=6$
superconformal $U(N)\times U(N)$
Chern-Simons-matter theory of Aharony,
Bergman, Jafferis and Maldacena (ABJM)
to the case when we study the scaling
limit of $N=6$ superconformal
$U(M)\times U(N)$ Chern-Simons-matter
theory of Aharony, Bergman and Jafferis
(ABJ). We show that if we extend the
ABJ theory in the same way as in
arXiv:0811.1540
 we can define a
 consistent limit that leads
 to $SU(N)$ L-BLG theory together
 with $U(M-N)_k$ Chern-Simons theory
 of level $k$.}

\def\tr{\mathrm{Tr}}

\def\tA{\tilde{A}}

\newcommand{\tX}{\tilde{X}}

\newcommand{\tchi}{\tilde{\chi}}

\newcommand{\tB}{\tilde{B}}

\newcommand{\tD}{\tilde{D}}

\def\\theta{\tilde{\psi}}

\newcommand{\mL}{\mathcal{L}}

\def \tY{\tilde{Y}}
\def\pb #1{\left\{#1\right\}}
\begin{document}
\section{Introduction and Summary}\label{first}
It is a long standing problem to find a
world-volume theory of $N$ M2-branes
that can be considered as a
generalization of the world-volume
theory of a single M2-brane
\cite{Bergshoeff:1987cm,Bergshoeff:1987qx}
\footnote{For review and extensive list
of references considering early years
of M2-brane theories, see
\cite{Dasgupta:2002iy,deWit:1999rh,Nicolai:1998ic}.}.
Following very nice analysis performed
in \cite{Basu:2004ed}, Bagger and
Lambert and Gustavson (BLG) formulated
a three-dimensional superconformal
Chern-Simons-matter theory that
successfully captures many aspects of
dynamics of $N$ M2-branes
\cite{Bagger:2006sk,Gustavsson:2007vu,Bagger:2007jr}.
The remarkable property of this theory
that it makes exceptional is that is
based on $3-$algebra. On the other hand
it was soon discovered that the
original formulation of BLG theory
describes only two coincident M2-branes
on condition that the $3-$algebra is
kept finite and it has a
positive-definite metric
\cite{VanRaamsdonk:2008ft,Bandres:2008vf,Papadopoulos:2008sk,
Gauntlett:2008uf,DeMedeiros:2008zm,deMedeiros:2008bf,Gomis:2008uv}.

In order to resolve this limitation it
was suggested in
\cite{Gomis:2008uv,Benvenuti:2008bt,Ho:2008ei,Bandres:2008kj,
Gomis:2008be} to use $3-$algebra with a
Lorentzian (indefinite) signature
metric. The resulting Lorentzian-BLG
(L-BLG) theory is $N=8$ superconformal
at the classical level even if its
interpretation as a quantum field
theory is still an open problem. In
particular, if one expand near a
classical vacuum that spontaneously
breaks the superconformal symmetry it
becomes equivalent
\cite{Ezhuthachan:2008ch,Mukhi:2008ux,Verlinde:2008di}
to a standard low-energy gauge theory
of multiple D2-branes that is
non-conformal $N=8$ supersymmetric $3d$
 $N=8$ SYM theory.

A different $3d$ superconformal
Chern-Simons-matter theory was proposed
by Aharony, Bergman, Jafferis and
Maldacena (ABJM)\cite{Aharony:2008ug}.
This theory possesses $N=6$
supersymmetry \cite{Benna:2008zy} and
it is interpreted as a theory that
describes $N$ coincident M2-branes at
the singularity of the orbifold
$\mathbf{C^4/Z_k} $. While the ABJM
theory also admits a $3-$algebra
interpretation \cite{Bagger:2008se} it
seems to be different from the original
L-BLG theory. More precisely, these
theories have different field content
and different symmetries.

An interesting suggestion how these two
theories are related was presented  in
\cite{Honma:2008jd} (see also
\cite{Honma:2008ef}), where it was
argued that
the L-BLG theory can be interpreted as
a certain limit of the ABJM theory in
which one sends the ABJM coupling $k$
(CS level) to infinity and at the same
time rescales some of the fields to
zero so that they decouple.

This  proposal was further clarified in
\cite{Antonyan:2008jf} where it was
argued that in order to relate these
two theories by scaling limit we have
to supplement the ABJM theory with an
extra ghost multiplet that is decoupled
from the ABJM fields. Further it was
explicitly shown that there exists a
limit of the $3-$algebra of ABJM theory
that is trivially extended by an extra
ghost generator that leads to the
Lorentzian $3-$algebra  of L-BLG
theory. It was also suggested there
that it is possible to interpret this
scaling limit as a definition of L-BLG
theory in terms of ABJM theory. In
particular, using this limit it can be
seen the relation between the L-BLG
theory and the $3d$ $N=8$ SYM theory
that describes $N$ D2-branes: Taking
the scaling limit and then giving one
of the scalars an expectation value is
equivalent to the procedure
\cite{Mukhi:2008ux,Distler:2008mk} for
obtaining the D2-brane theory from the
ABJM theory.

Further more general forms of
 ABJM theory  were
introduced  in \cite{Aharony:2008gk}.
One  example of such a more general
theory is  $U(M)_k \times U(N)_{-k}$
Chern-Simons-matter that has the same
matter content and interactions as in
\cite{Aharony:2008ug} but with $M\neq
N$.  From the point of view of
M2-branes as probes of
$\mathbf{C^4/Z_k}$ singularity these
theories arise (for $M>N$) when we have
$(M-N)$ fractional M2-branes that are
localized at the singularity together
with $N$ M2-branes that are free to
move around.  It was argued in
\cite{Aharony:2008gk} that these
theories exist as  well defined quantum
field theories in case when $|M-N|\leq
k$ and that in this case the
gravitational dual description is
$AdS_4\times \mathbf{S^7/Z_k}$
background as in \cite{Aharony:2008ug}
but with an additional "torsion flux"
that takes values in
$H^4(\mathbf{S^7}/\mathbf{Z_k},\mathbf{Z})=\mathbf{Z_k}$.

Since  the ABJ theory
 \cite{Aharony:2008ug} is more general
 than the  ABJM theory
 we mean that it deserves
to be studied further. In particular,
we would like to see how to define the
scaling limit similar to the limit
given in \cite{Honma:2008jd} in case of
the ABJ theory.
 In this note
 we introduce such a scaling limit and
show that it leads to well defined
theory. Concretely, we implement such a
scaling limit that corresponds to the
decoupling of all massive degrees of
freedom when we move $N$ M2-branes far
away
 from the origin
of $\mathbf{C^4/Z_k}$. According to
general arguments \cite{Aharony:2008ug}
we can expect that  the low energy
modes are pure $U(M-N)_k$ Chern-Simons
theory that describes dynamics of $N-M$
fractional M2-branes localized at the
origin of $\mathbf{C^4/Z_k}$ together
with   L-BLG theory that describes
dynamics of $N$ M2-branes in flat
space.  In fact, we will study the
spectrum of fluctuation modes around
the vacuum state that corresponds to
$N$ M2-branes moving from the origin
and we identify spectrum of massless
and massive modes that agrees with
observation given in
\cite{Aharony:2008ug}. Then we define
such a scaling limit that decouples
these massive modes and retains the
massless ones together with auxiliary
fields that are well known from L-BLG
theory. We explicitly show, following
\cite{Honma:2008jd} and
\cite{Antonyan:2008jf}  that this limit
leads to $U(N-M)_k$ CS theory together
with L-BLG theory and that these two
theories are decoupled.

This result implies that L-BLG theory
can be defined from ABJ theory as well
in the limit when we appropriately
redefine  the fields and the level $k$,
add ghosts fields
\cite{Antonyan:2008jf} and then send
the small parameter to zero.

The extension of this work is as
follows. We can ask the question how to
define the decoupling limit and what is
the resulting theory in case of
$(U(N)\times U(N))^n$ superconformal
quiver gauge theories
\cite{Benna:2008zy} (see also
\cite{Hashimoto:2008ij,
Terashima:2008ba,Hosomichi:2008jb,Hosomichi:2008jd}).
We hope to return to this problem in
future.

The organization of this paper is as
follows. In the next section
(\ref{second}) we review basic facts
considering ABJ theory. Then in section
(\ref{third}) we analyze the situation
when $N$ M2-branes is localized far
away from the origin of $\mathbf{C^4}
/\mathbf{Z_k}$. We determine massless
and massive modes that propagate around
this vacuum solution. Using this result
we introduce in section (\ref{fourth})
the scaling limit in ABJ theory that
decouple the massive and massless modes
and leads to $U(N-M)_k$ CS theory of
level $k$ together with $SU(N)$ L-BLG
theory.

\section{Aspect of ABJ
Theory}\label{second} The difference
between ABJM and ABJ theory is that
fractional M2-branes are added as a new
parameter in dual theory. In other
words theory possesses three parameters
$M,N,k$ that all are integer valued.
Now we  briefly  review basic facts
considering this theory
\begin{itemize}
\item Let us consider M2-brane as a
probe of $\mathbf{C^4/Z_k}$
singularity. As was argued in
\cite{Aharony:2008gk} the ABJ theories
arise when, in addition to $N$
M2-branes that can move around
$\mathbf{C^4/Z_k}$ singularity, $(N-M)$
fractional M2-branes that are localized
at orbifold singularity.
\item The classical field theory
description of this theory is given by
$N=6$ superconformal theory with gauge
group $U(M) \times U(N)$ where $M\neq
N$. On the other hand it was argued in
\cite{Aharony:2008gk}
 that these theories exist as a
unitary superconformal theories only
for $|M-N|\leq k$. \item The gravity
dual of these unitary theories is
$AdS_4\times \mathbf{S^7/Z_k}$
background that was originally
introduced in \cite{Aharony:2008ug} but
now with additional "torsion flux" that
takes values in
$H^4(\mathbf{S^7}/\mathbf{Z_k},
\mathbf{Z})=\mathbf{Z_k}$. In fact,
this is a good description  of the
gravitational dual when $N\gg k^5$. In
case when $k\ll N \ll k^5$ the
appropriate description is in terms of
type IIA string theory on $AdS_4\times
CP_3$ with a discrete holonomy of the
NSNS 2-form field in the $CP^1\subset
CP^3$.
\end{itemize}
Let us be more concrete in description
of ABJ theory. This theory is $N=6$
 supersymmetric Chern-Simons-matter theory
with  two gauge groups or ranks $M,N$ and
levels $k$ and $-k$ respectively.
Further, this theory is characterized
by following properties:
\begin{itemize}
\item Gauge and global symmetries:
\vskip 4mm \quad gauge symmetry: \
$U(M)\otimes U(N) $ \vskip 4mm \quad
global symmetry: \ $SU(4)$
\item
The field content of given theory is as
follows:
\begin{eqnarray}
& &A_\mu^{(L)}: \ \mathrm{Adj}(U(M)) \
, \quad  A_\mu^{(R)}: \ \mathrm{Adj}(U(N))
 \ . \nonumber \\
\end{eqnarray}
Further we have $M\times N$ matrix
valued matter fields-$4$ complex scalar
$Y^A (A=1,2,3,4)$ and their hermitian
conjugates $Y^\dag_A$.
 We have also $M\times
N$ matrix valued  fermions $\psi_A$
 together with their hermitian
conjugates $\psi^{A\dag}$.
 Fields with
raised $A$ index transform in the
$\mathbf{4}$ of $R$ symmetry $SU(4)$
group and those with lowered index
transform in the
$\overline{\mathbf{4}}$
representations.
\end{itemize}
%
The corresponding Lagrangian has the
following form
\begin{eqnarray}\label{Act}
\mL &=&-\tr (D^\mu Y^\dag_A D_\mu
Y^A)-i \tr (\overline{\psi}^{A\dag}
\gamma^\mu D_\mu\psi_A)-V+\mL_{CS}-
\nonumber \\
&-& i\frac{2\pi}{k} \tr
(\overline{\psi}^{A\dag}\psi_A
 Y_B^\dag
 Y^B-\overline{\psi}^{A\dag}Y^B
 Y^{\dag}_B \psi_A)
 +2\frac{2\pi}{k}
 \tr (\overline{\psi}^{A\dag}\psi_B
 Y^\dag_A
 Y^B-\overline{\psi}^{A\dag}Y^B
 Y_A^\dag \psi_B)+\nonumber \\
 &+& i\frac{2\pi}{k}
 \epsilon_{ABCD} \tr
 (\overline{\psi}^{A\dag}Y^C
 \psi^{B\dag}Y^D)-i\frac{2\pi}{k}
 \epsilon^{ABCD}\tr(Y^\dag_D
 \overline{\psi}_A Y^\dag_C \psi_B) \ ,
  \nonumber \\
 \end{eqnarray}
where $\mL_{CS}$ is a Chern-Simons term
and $V(Y)$ is a sextic scalar potential
\begin{eqnarray}\label{CSact}
\mL_{CS}&=&\frac{k}{4\pi}
\epsilon^{\mu\nu\lambda}
\tr[A^{(L)}_\mu
\partial_\nu A_\lambda^{(L)}+
\frac{2i}{3}A_\mu^{(L)}A_\nu^{(L)}A_\lambda^{(L)}]
-\nonumber \\
&-&\frac{k}{4\pi}
\epsilon^{\mu\nu\lambda}
\tr[A_\mu^{(R)}\partial_\nu
A_\lambda^{(R)}+\frac{2i}{3}A_\mu^{(R)}
A_\nu^{(R)}A_\lambda^{(R)}] \ ,
\nonumber \\
V(Y)&=& -\frac{4\pi^2}{3k^2}\tr [
Y^AY_A^\dag Y^B Y_B^\dag Y^C Y_C^\dag
+Y_A^\dag Y^A Y_B^\dag Y^B Y_C^\dag
Y^C+\nonumber \\
&+& 4 Y^AY_B^\dag Y^C Y_A^\dag Y^B
Y_C^\dag -6 Y^A Y_B^\dag Y^B Y_A^\dag
Y^C Y_C^\dag] \ . \nonumber \\
\end{eqnarray}
Further,   the covariant derivatives
are defined as
\begin{eqnarray}\label{Dcov}
D_\mu Y^A=\partial_\mu Y^A
+iA^{(L)}_\mu Y^A-iY^A
A^{(R)}_\mu \ , \quad
D_\mu Y^\dag_A=
\partial_\mu Y^\dag_A -iY^\dag_A
A_\mu^{(L)}+i A_\mu^{(R)}Y^\dag_A
 \  \nonumber \\
\end{eqnarray}
and for fermions
\begin{eqnarray}
D_\mu \psi_A=
\partial_\mu \psi_A+i
A^{(L)}_\mu \psi_A-i\psi_A A_\mu^{(R)}
\ , \quad D_\mu \psi^{A\dag }=
\partial_\mu \psi^{A\dag}
-i\psi^{A\dag }A_\mu^{(L)}+ i
A_\mu^{(R)}\psi^{A\dag} \ .  \nonumber \\
\end{eqnarray}
After review of basic properties of ABJ
theory we will  analyze the spectrum of
 fluctuation modes
around the configuration when $N$
M2-branes are localized far away from
the origin of $\mathbf{C^4/Z_k}$.
%
%
\section{$N$ M2-branes Displaced From
The Origin of $\mathbf{C^4/Z_k}$}\label{third}
 In this section we study
 solutions of the $U(M)\times U(N)$
theory that describes situation when we
move $N$ M2-branes from the origin of
$\mathbf{C^4/Z_k}$. For concreteness we
presume that  $M>N$. Then it is natural
to write  $A_\mu^{(L)}$ in the  form
\begin{eqnarray}
A_\mu^{(L)}=\left(\begin{array}{cc}
A_{11 \mu}^{(L)} & A_{12\mu}^{(L)} \\
A_{21 \mu}^{(L)} & A_{22\mu}^{(L)} \\
\end{array}\right) \ ,
\end{eqnarray}
where $A_{11\mu}^{(L)}$ is $(N-M)\times
(N-M)$ matrix, $A_{12\mu}^{(L)}$ are
$(M-N)\times N$ and $A_{21\mu}^{(L)}$
 $N\times (M-N)$ matrices. Finally
$A_{22\mu}^{(L)}$ is $N\times N$
matrix. In the same way we write
\begin{equation}
Y^A=\left(\begin{array}{cc}
Z^A \\
Y^A_0 \mathrm{I}_{N\times N}+\tY^A
\\ \end{array}\right)
\end{equation}
where $Z^A$ is $(M-N)\times N$ matrix and
$\tY^A$ is $N\times N$ matrix with $\tr \tY^A=0$.

We are interested in configuration when
we separate $N$ M2-branes far away from
the  origin of $\mathbf{C^4/Z_k}$. In
other words we consider the solution of
the equation of motion of ABJ theory in
the form
\begin{equation}\label{Yvac}
Y^A=\left(\begin{array}{cc} 0_{M\times
N} \\
R^A \mathrm{I}_{N\times N} \\ \end{array}\right)
\end{equation}
and  where all other  fields are equal
to zero.
Then in  order to find the spectrum of
fluctuations  we expand the scalar
fields around the ansatz (\ref{Yvac})
as
\begin{equation}
Y^A=\left(\begin{array}{cc} Z^A \\
(R^A+ Y_0^A) \mathrm{I}_{N\times N}+\tY^A \\
\end{array}\right) \ , \quad \tr \tY^A=0 \ .
\end{equation}
Inserting this ansatz into the definition
of the covariant derivative
(\ref{Dcov}) we obtain
\begin{eqnarray}
& &\partial_\mu Y^A+ i A_\mu^{(L)}Y^A-
iY^A A_\mu^{(R)}=\nonumber \\
&=&\left(\begin{array}{cc}
\partial_\mu Z^A+iA_{11\mu}^{(L)}Z^A-
iZ^A A_\mu -i\frac{1}{2}Z^A
B_\mu+iA_{12\mu}^{(L)}(  Y_0^A
I_{N\times N}
+\tY^A) \\
 \partial_\mu Y_0^A I_{N\times N}+
\partial_\mu \tY^A+
i\left[A_\mu,\tY^A\right]- iY_0^A B_\mu
-i\frac{1}{2}\pb{B_\mu,\tY^A}
 \\ \end{array}\right)
+\left(\begin{array}{cc}i A^{(L)}_{12\mu }R \\
-iB_\mu R\\
\end{array}\right) \ ,
\nonumber \\
\end{eqnarray}
where we introduced  fields $B_\mu$ and
$A_\mu$ as a combinations of
$A_{22\mu}^{(L)}$ and $A_\mu^R$
\begin{equation}\label{AB}
A_{22\mu}^{(L)}=A_\mu- \frac{1}{2}B_\mu
\ , \quad A_{\mu}^{(R)}= A_\mu+
\frac{1}{2}B_\mu
\end{equation}
This result implies that due to the
Highs mechanism fields $A^{(L)}_{12\mu}$
and $B_\mu$ become massive with mass terms
equal to
\begin{equation}
\tr (A^{(L)}_{12\mu}A^{(L)\mu}_{21})R^2=
\tr (A^{(L)}_{12\mu}(A^{(L)\mu})^\dag_{12})R^2
 \ ,
 \quad
 \tr (B_\mu B^\mu) R^2 \ ,
\end{equation}
where $R^2=R_AR^A$ and where we used
the fact that $(A_{12\mu}^{(L)})^\dag=
A_{21\mu}^{(L)}$. We see that these
fields become infinite massive in the
limit $R\rightarrow \infty$.  However
there is an important difference
between $B_\mu$ and $A_{12\mu}^{(L)}$
since  $B_\mu$ is auxiliary field while
$A_{12\mu}^{(L)}$ is massive vector
field with ordinary kinetic term. To
see this note that introducing the
variables (\ref{AB}) the Chern-Simons
Lagrangian density (\ref{CSact}) can be
rewritten into the form
\begin{eqnarray}
\mL_{CS}&=&
\frac{k}{2\pi}\epsilon^{\mu\nu\lambda}
\tr \left[A_{11\mu}^{(L)}\partial_\nu
A^{(L)}_{11\lambda} +\frac{2i}{3}
A_{11\mu}^{(L)}A_{11\nu}^{(L)}
A_{11\lambda}^{(L)}+\right.\nonumber \\
&+& 2A_{12\mu}^{(L)}\partial_\nu A_{21
\lambda}^{(L)} +2i (A_{11\mu}^{(L)}
A_{12\nu}^{(L)} A_{21\lambda}^{(L)}+
A_{22\mu}^{(L)}
A_{21\nu}^{(L)}A_{12\lambda}^{(L)})
-\nonumber \\
&-&\left. B_\mu(\partial_\nu A_\lambda-
\partial_\lambda
A_\nu+i[A_\nu,A_\lambda]) -\frac{i}{6}
B_\mu B_\nu B_\lambda\right] \nonumber \\
\end{eqnarray}
that shows that there is no kinetic
terms for $B_\mu$ that confirms the
claim that $B_\mu$ is auxiliary.

Let us now consider the fluctuation
modes $Z^A,\tY^A$ and $Y_0^A$. It is
easy to see from  the form of the
scalar potential $V_B$ that the fields
$Z^A$ are massive with mass
proportional $R^4$ while $Y_0^A,\tY^A$
are massless.

In the same way we can proceed in case of
fermions. Explicitly, we  write
bi-fundamental
 fermions as
\begin{equation}
\psi_A=\left(\begin{array}{cc} \chi_A
\\
\psi_A^0\mathrm{I}_{N\times N}+\theta_A \\
\end{array}\right) \ , \quad  \tr \theta_A=0 \
.
\end{equation}
Then we can straightforwardly analyze
the scalar part of the Lagrangian
(\ref{Act}) and determine that the
modes $\chi_A$ become massive with mass
proportional to $R^2$
 while the fields $\psi_A^0$ and  $\theta_A$
are  massless.

Let us now  summarize the field content
around the configuration with large
$\left<Y^AY_A^\dag\right>= R^2$
corresponding
 separation of $N$
M2-branes from the origin of $\mathbf{C^4/Z_k}$:
\begin{eqnarray}\label{ficon}
& & A^{(L)}_{11\mu} \ , \quad  A_{\mu} \ , \quad
\psi_A^0 \ , \quad  \theta_A \ , \quad  Y_0^A \ , \quad
\tY^A: \quad
\mathrm{massless} \ , \nonumber \\
& & Z^A \ , \quad  A_{12\mu}^{(L)} \ , \quad
 \chi_A: \quad  \mathrm{massive} \ , \nonumber \\
 & & B_\mu: \quad \mathrm{auxiliary} \ .
 \nonumber \\
\end{eqnarray}
Since the scaling limit defined in
\cite{Honma:2008jd} can be interpreted
as a limit when we move the M2-branes
infinitely far from the singularity
together with sending $k$ to infinity
we can expect that similar limit exists
in the ABJ theory as well.
\section{Scaling limit}\label{fourth}
Motivated by the analysis performed in
previous section we would like to
define the scaling limit that decouple
massive fields given in (\ref{ficon})
 and leads to a
theory of massless fields together with
auxiliary fields only. To do this we
propose the scaling limit
 in the form
\begin{eqnarray}\label{scalgauge}
A_{11\mu}^{(L)}&=&A_{11\mu}^{(L)}\ , \quad
A_{12\mu}^{(L)}=\epsilon^2 \tA_{12\mu}
\ , \quad  A_{21\mu}^{(L)}= \epsilon^2
\tA_{21\mu}
\ , \nonumber \\
A^{(L)}_{11\mu}&=&
A_\mu-\frac{1}{2}\epsilon B_\mu \ ,
\quad A^{(R)}_\mu= A_\mu+
\frac{1}{2}\epsilon B_\mu \ , \nonumber
\\
\end{eqnarray}
where $\epsilon$ is small parameter
that controls the scaling limit and
where we  take $\epsilon\rightarrow 0$
in the end.
 Note also that $A_\mu$ and $B_\mu$
belong to the algebra of
$\mathbf{u}(N)$.

To begin with we insert redefined gauge
fields (\ref{scalgauge}) into the
Chern-Simons Lagrangian and  we obtain
\begin{eqnarray}
\mL_{CS}
&=&\frac{k}{2\pi}
\epsilon^{\mu\nu\lambda} \tr
(A_{11\mu}^{(L)}\partial_\nu
A_{11\lambda}^{(L)}+
\frac{2i}{3}A^{(L)}_{11\mu}
A^{(L)}_{11\nu}A^{(L)}_{11\lambda}
\nonumber \\
&+&2i\epsilon^4
A_{11\mu}^{(L)}A_{12\nu}^{(L)}A_{21\lambda}^{(L)}
+2i \epsilon^4
A_{22\mu}^{(L)}A_{21\nu}^{(L)}A_{12\lambda}^{(L)})+
\nonumber \\
&+&\frac{k}{2\pi}
\epsilon^{\mu\nu\lambda} [-\epsilon
B_\mu (\partial_\nu
A_\lambda-\partial_\lambda A_\mu+i
[A_\nu, A_\lambda])
+O(\epsilon^2)]\equiv
\nonumber \\
&\equiv& \mL^{(1)}+\mL^{(2)} \ , \nonumber \\
\end{eqnarray}
where
\begin{eqnarray}
\mL^{(1)}&=&\frac{k}{2\pi}\epsilon^{\mu\nu\lambda}
 \tr
(A_{11\mu}^{(L)}\partial_\nu
A_{11\rho}^{(L)}+
\frac{2i}{3}A^{(L)}_{11\mu}
A^{(L)}_{11\nu}A^{(L)}_{11\lambda}) \ ,
\nonumber \\
\mL^{(2)}&=&-\frac{k\epsilon}{2\pi}
\epsilon^{\mu\nu\lambda} \tr B_\mu
(\partial_\nu A_\rho-\partial_\lambda
A_\mu+i [A_\nu, A_\lambda]) \ .  \nonumber \\
\end{eqnarray}
We see that in order to decouple the
massive states we should keep $k$
unscaled in the first part of the
Lagrangian $\mL^{(1)}$ while in the
second one we should take
$k=\frac{1}{\epsilon}\tilde{k}$. Then
in the limit $\epsilon \rightarrow 0$
we end with
\begin{eqnarray}
\mL^{(1)}&=&\frac{k}{2\pi}
\epsilon^{\mu\nu\lambda} \tr
(A_{11\mu}^{(L)}\partial_\nu
A_{11\lambda}^{(L)}+
\frac{2i}{3}A^{(L)}_{11\mu}
A^{(L)}_{11\nu}A^{(L)}_{11\lambda}) \ ,
\nonumber
\\
\mL^{(2)}&=&- \frac{\tilde{k}}{2\pi}
\epsilon^{\mu\nu\lambda} \tr B_\mu
F_{\nu\lambda} \ , \quad
F_{\nu\lambda}=
\partial_\nu A_\lambda-\partial_\lambda
A_\mu+i [A_\nu, A_\lambda] \ . \nonumber \\
\end{eqnarray}
 Let us now
give the physical interpretation of the
result above. The Lagrangian density
$\mL^{(1)}$ describes $U(N-M)_k$
Chern-Simons theory living on the
world-volume of fractional M2-branes
that are localized at the origin of
$\mathbf{C^4/Z_k}$. Considering the
Lagrangian density $\mL^{(2)}$ we
should split $B_\mu $ and $A_\mu$ gauge
fields into $U(1)$ and $SU(N)$ parts as
\begin{eqnarray}\label{SU}
A_\mu=A_\mu^0\mathrm{I}_{N\times N}+\tA_\mu
 \ , \quad   \tr \tA_\mu=0 \ , \nonumber \\
\tilde{k}B_\mu= \epsilon^2 B_\mu^0
\mathrm{I}_{N\times N}+ \tB_\mu \ , \quad  \tr
\tB_\mu=0 \ , \nonumber
\\
\end{eqnarray}
where
 for letter convenience we rescaled
the $U(1)$ part of $B$ field with
$\epsilon^2$.
 Then $\mL^{(2)}$ takes the form
\begin{equation}
\mL^{(2)}=-\frac{\epsilon^2 N}{2\pi}
\epsilon^{\mu\nu\lambda} B^0_\mu
F^0_{\nu\lambda} - \frac{1}{2\pi}
\epsilon^{\mu\nu\lambda}\tr\tB_\mu
\tilde{F}_{\nu\lambda}= -
\frac{1}{2\pi}
\epsilon^{\mu\nu\lambda}\tr\tB_\mu
\tilde{F}_{\nu\lambda} \ .
\end{equation}
that is precisely the gauge part of
L-BLG theory.

 Now we consider  the
scaling limit  of matter fields
$Y^A,\psi_A$. Following analysis
presented in  previous section we
suggest that they scale as
\begin{eqnarray}\label{scalscalar}
Y^A=\left(\begin{array}{cc}
\epsilon^2 Z^A \\
\frac{1}{\epsilon}Y^A_+ \mathrm{I}_{N\times
N}+\tY^A
\\ \end{array}\right) \ , \quad \tr \tY^A=0
\nonumber \\
\psi_A= \left(\begin{array}{cc}
\epsilon \tchi_A \\
\frac{1}{\epsilon}\psi_{A+} \mathrm{I}_{N\times
N}+\theta_A \end{array}\right) \ ,
\quad \tr \theta_A=0\ .  \nonumber
\\
\end{eqnarray}
Using the definition of the covariant
derivative (\ref{Dcov}) we  obtain
\begin{eqnarray}
D_\mu Y^A &=&
\left(\begin{array}{cc} \epsilon^2
\partial_\mu Z^A+ i\epsilon^2 A_{11\mu}^{(L)}Z^A- Z
i\epsilon^2 A^{(R)}_\mu+ i \epsilon^2
A_{12\mu}^{(L)}(\frac{1}{\epsilon}Y_+^A
\mathrm{I}_{N\times N}+\tY^A) \\
\frac{1}{\epsilon}\partial_\mu Y_+^A
\mathrm{I}_{N\times N}+
\partial_\mu \tY^A+
i\left[A_\mu,\tY^A\right]- i Y_+^A
B_\mu
-\frac{i}{2}\epsilon\pb{B_\mu,\tY^A}
 \\ \end{array}\right)
 \rightarrow
\nonumber \\
&=&\left(\begin{array}{cc}  O(\epsilon)_{(M-N)\times N}\\
\frac{1}{\epsilon}\partial_\mu Y_+^A
\mathrm{I}_{N\times N}+
\partial_\mu \tY^A+
i\left[A_\mu,\tY^A\right]- iY_+^A
(\epsilon^2 B^0_\mu \mathrm{I}_{N\times
N}+ \tB_\mu)
-\frac{i}{2}\epsilon\pb{B_\mu ,\tY^A}
 \\ \end{array}\right)\equiv \nonumber
 \\
& &\left(\begin{array}{cc} O(\epsilon)_{(M-N)\times N} \\
\frac{1}{\epsilon}\partial_\mu Y_+^A
\mathrm{I}_{N\times N}+ \tD_\mu \tY^A-
\frac{i}{2}\epsilon\pb{\tB_\mu,\tY^A}
 \\ \end{array}\right) \ ,
\nonumber \\
 \end{eqnarray}
where we defined
\begin{equation}
\tD_\mu \tY^A=
\partial_\mu \tY^A+i[
A_\mu,\tY^A]-iY_+^A \tB_\mu \ .
\end{equation}
In the same way we find that
\begin{eqnarray}
D_\mu Y^\dag_A&=&
\left(\begin{array}{cc}
O(\epsilon)_{N\times (M-N)} &  \quad
 \frac{1}{\epsilon}
\partial_\mu Y_{+A}^\dag \mathrm{I}_{N\times N}+
(\tilde{D}_\mu \tY_A)^\dag
+
\frac{i\epsilon}{2}\pb{\tB_\mu,\tY_A^\dag}\end{array}
\right)
 \nonumber \\
 \end{eqnarray}
where
\begin{equation}
(\tilde{D}_\mu Y_A)^\dag=
\partial_\mu \tY_A^\dag
-i[\tY_A^\dag,A_\mu] + iY_{A+}^\dag
\tB_\mu \ .
\end{equation}
Then using these results we find that
the kinetic term for $Y_A$ takes the
form
\begin{eqnarray}\label{kintermsc}
\tr (D_\mu Y^\dag_A D^\mu Y^A )=
\frac{N}{\epsilon^2} \partial_\mu
Y_{A+}^\dag
\partial^\mu Y^A_++
\tr \tD_\mu \tY^\dag_A \tD^\mu \tY^A-
\nonumber \\
-i\partial_\mu Y^\dag_{+A}\tr
(\tB^{\mu}\tY^A) +i\partial_\mu Y^A_+
\tr (\tB^{\dag\mu}
\tY_A^\dag) \ .  \nonumber \\
\end{eqnarray}
We see that the first term diverges in
the limit $\epsilon\rightarrow 0$ and
hence the scaling limit in the form, as
was presented in   \cite{Honma:2008jd}
seems to be not complete. The careful
discussion of  this issue and its
resolution was given in
\cite{Antonyan:2008jf} and we recommend
this paper for more details. The result
of the analysis presented there   is
that in order to have well defined
scaling limit we have to  add an extra
term to the bosonic ABJ Lagrangian
\begin{equation}\label{addional}
N \partial_\mu U_A^\dag \partial^\mu
U^A \ .
\end{equation}
Note that  (\ref{addional}) has a
"wrong" sing of the kinetic term and
hence $U^A$ can be interpreted as an
extra ghost. In fact, following
arguments given  in
\cite{Antonyan:2008jf}  it is natural
to extend the original ABJ action by
this "ghost" term since it would be
puzzling that we can derive L-BLG
action that has an indefinite
kinetic-term signature from a
manifestly definite ABJ action by a
regular scaling limit. It is also
important to stress that at the level
of ABJ theory the extra ghost term is
decoupled. On the other hand it gets
effectively coupled through the
following redefinition
\begin{equation}\label{redU}
U^A=-\frac{1}{\epsilon}Y_+^A+
\epsilon\frac{1}{N}Y^A_- \
\end{equation}
in the process when we implement the
scaling limit. Note that through this
redefinition we introduced new scalar
field $Y^A_-$ that plays crucial role
in L-BLG theory. Then, using
(\ref{redU}) we obtain that
(\ref{kintermsc}) together with
(\ref{addional}) give finite
contribution to the action in the limit
$\epsilon\rightarrow 0$
\begin{eqnarray}\label{kinfinal}
& & N \partial_\mu U^\dag \partial^\mu
U^A -\frac{N}{\epsilon^2} \partial_\mu
Y_{A+}
\partial^\mu Y^A_+-
\tr \tD_\mu \tY^\dag_A \tD^\mu \tY^A+
\nonumber \\
&+ &i\partial_\mu Y_{+A}\tr
(B^{\mu}\tY^A) -i\partial_\mu Y^A_+ \tr
(B^{\dag\mu}
\tY_A^\dag)= \nonumber \\
&=&-\partial_\mu Y_{+}^A\partial^\mu
Y_{-A}^{\dag}-
\partial_\mu Y_{+ A}^\dag
\partial^\mu Y_{-}^A-
\tr \tD_\mu \tY^\dag_A \tD^\mu \tY^A+
\nonumber \\
&+& i\partial_\mu Y_{+A}^\dag\tr
(B^{\mu}\tY^A) -i\partial_\mu Y^A_+ \tr
(B^{\dag\mu} \tY_A^\dag)=
\nonumber \\
&=&-2\partial_\mu X^I_+\partial^\mu \tX^I_-
-2\partial_\mu \tX^I_+ \tr (B^\mu \tX^I)-
\nonumber \\
&-&(D_\mu \tX^I-X^I_+B_\mu)
\eta^{\mu\nu}
(D_\nu \tX^I-X^I_+B_\nu) \nonumber \\
\end{eqnarray}
using the relations  between
real scalar fields
$X^I_\pm \ ,  X^{I*}_\pm=X^I_\pm \ ,
\quad \tX^{I\dag}=\tX^I,I=1,\dots 8 $ and
complexified scalars $Y^A$:
\begin{eqnarray}
Y^A_\pm=X^{2A-1}_\pm+iX^{2A}_\pm \ ,
\quad
Y^{\dag}_{A\pm}=X^{2A-1}_\pm-iX^{2A}_\pm
\ ,
\nonumber \\
\tY^A=-\tX^{2A}+i\tX^{2A-1} \ , \quad
\tY_A^\dag=-\tX^{2A}-i\tX^{2A-1} \ ,
\end{eqnarray}
and where
\begin{equation}
D_\mu \tX^I=\partial_\mu \tX^I+
i[A_\mu,\tX^I] \ .
\end{equation}
Note that (\ref{kinfinal})  takes
precisely the same form as the bosonic
kinetic term in L-BLG theory, up to trivial
rescaling of scalar fields.
%
Let us now consider the scaling limit
of  the kinetic term for fermions. If
we insert (\ref{scalgauge}),(
\ref{scalscalar}) and (\ref{SU}) into
it and then take the limit
 $\epsilon \rightarrow 0$ we obtain
\begin{eqnarray}\label{kinf}
i\tr (\overline{\psi}^{A\dag}
\gamma^\mu D_\mu\psi_A)&=& iN\frac{1}{\epsilon^2}
\overline{\psi}^{A\dag}_{+}\gamma^\mu
\partial_\mu \psi_{A+}+
\nonumber \\
\nonumber \\
&+& i\tr (\overline{\theta}^{\dag
A}\gamma^\mu \tilde{D}_\mu \theta_A)- i
\overline{ \psi}^{A\dag}_+ \gamma^\mu
\tr \pb{\tB_\mu,\theta_A} \ .
\nonumber \\
 \end{eqnarray}
In the same way as in the analysis of
the bosonic kinetic term
 we add to the Lagrangian the ghost
contribution
\begin{eqnarray}\label{ghostf}
-iN\overline{V}^{\dag A}
\gamma^\mu\partial_\mu V_A \ ,
\nonumber \\
\end{eqnarray}
where now $V^\mu$ is fermionic field.
Then if we perform following
redefinition
\begin{equation}
V_A=-\frac{1}{\epsilon}\psi_{+A}
+\frac{\epsilon}{N}\psi_{-A}
\ ,
\end{equation}
where we introduced new fermion field
$\psi_{-A}$ we find that the kinetic
term for fermions (\ref{kinf}) together
with (\ref{ghostf}) are well defined
even in the limit $\epsilon\rightarrow
0$
\begin{eqnarray}
i\tr
\overline{\theta}^{A\dag}\gamma^\mu
\tilde{D}_\mu \theta_A- i
\overline{\psi}_+^A \gamma^\mu \tr
\pb{\tB_\mu,\theta_A}
+i\overline{\psi}_+^{A\dag}
\gamma^\mu\partial_\mu \psi_{-A}+
i\overline{\psi}_-^{A\dag}
\gamma^\mu\partial_\mu \psi_{+A} \ .
\nonumber \\
\end{eqnarray}
Then it is easy to see that this
  final expression
can be  rewritten into the manifestly
$SO(8)$ invariant fermionic kinetic
term of L-BLG model.

Let us now consider the scaling limit
in the scalar  terms in (\ref{Act}) and
(\ref{CSact}). In fact, the analysis
will be almost the same as in
\cite{Honma:2008jd} with difference
that we have to explicit show that the
modes $Z^A$ decouple in the scaling
limit. For example, let us consider
following contribution to the bosonic
potential (\ref{CSact})
\begin{equation}
\frac{1}{k^2}\tr (Y^BY_B^\dag Y^C
Y_C^\dag Y^AY_A^\dag) \ .
\end{equation}
Then using the scaling limit of scalar
fields (\ref{scalscalar}) we obtain
\begin{eqnarray}
\frac{1}{k^2}\tr (Y^BY_B^\dag Y^C
Y_C^\dag
Y^AY_A^\dag)=\nonumber \\
=\frac{\epsilon^4}{\tilde{k}^2}
\tr (\left(\begin{array}{cc} \epsilon^4
Z^BZ_B & \epsilon
Z^B\tY_B^\dag+\epsilon^2
Z^B\tY_B^\dag \\
\epsilon Y^B_+Z_B^\dag+\epsilon^2
Y^\dag_B Z^B & \frac{1}{\epsilon^2}
Y_+^B\tY^\dag_{+B}
+\frac{1}{\epsilon}(Y_+^B\tY_B^\dag+
Y_{+B}^\dag \tY^B)+ \tY^B\tY_B^\dag
\\
\end{array}\right)^3=\nonumber \\
=\frac{1}{\epsilon^2\tilde{k}^2} \tr (
Y_+^A Y^\dag_{+A}
+\epsilon(Y_+^A\tY_A^\dag+ Y_{+A}^\dag
\tY^A)+\epsilon^2 \tY^A\tY_A^\dag)^3
+O(\epsilon^4)
\nonumber \\
\end{eqnarray}
and we see that the modes $Z^A$ really
decouple. Further, the final expression
takes exactly the same form as the
contribution to the potential of
$U(N)\times U(N)$ ABJM theory. However
the analysis of this potential was
performed in \cite{Honma:2008jd} with
the following results. Due to the
decomposition of the $Y^A$ modes into
trace part $Y^A_+$ and traceless parts
$\tY^A$ and using the fact that the
potential is sextic we find that the
potential is sum of $V_B=\sum_{n=0}^6
V_B^{(n)}$ where $V_B^{(n)}$ contains
$n$ $Y_+$ fields and $(6-n)$ $\tY$
fields. Further, using the fact that
the potential is multiplied with
$\frac{1}{k^2}$ we obtain that
$V_B^{(n)}$ term scales as
$\epsilon^{2-n}$ in the limit
$\epsilon\rightarrow 0$.Then it can be
shown that the terms $V_B^{(n)}$ vanish
for $n>3$. On the other hand the
potential terms with $n<2$ vanish in
the limit $\epsilon\rightarrow 0$ and
the non-zero contribution comes from
$V_B^{(2)}$ part of the potential.
 Then
it was further shown in
\cite{Honma:2008jd} that this potential
has full $SO(8)$ symmetry and finally
that this potential exactly reproduces
potential of L-BLG theory.

In the same way we can analyze the
scalar terms in (\ref{Act})
 that contain both
fermions and bosons. Let us consider
for example an expression
\begin{equation}\label{FS}
\frac{2\pi}{k}\tr
(\overline{\psi}^{\dag A}\psi_A
Y_B^\dag Y^B) \ .
\end{equation}
Then using the scaling
(\ref{scalscalar}) and
$k=\frac{1}{\epsilon}\tilde{k}$ we find
that in  the limit $\epsilon\rightarrow
0$ (\ref{FS}) reduces into
\begin{eqnarray}
& &\frac{2\pi}{k}\tr
(\overline{\psi}^{\dag A}\psi_A
Y_B^\dag Y^B)\rightarrow
\frac{2\pi}{\tilde{k}\epsilon^3} \tr
(\overline{\psi}_+^{\dag A}I_{N\times
N}+\epsilon \overline{\theta}^{\dag A})
\times \nonumber \\
&\times &(\psi_{B+}I_{N\times N}+
\epsilon \theta_B) \times
(Y_{+B}I_{N\times N}+ \epsilon\tY_B)
(Y^B_{+}I_{N\times N}+\epsilon\tY^B)
\nonumber
\\
\end{eqnarray}
that has exactly the same form as in
$U(N)\times U(N)$ ABJM theory and
consequently the analysis performed in
\cite{Honma:2008jd} can be applied for
this case as well.

In summary, we have found the scaling
limit of ABJ theory defined by
(\ref{scalgauge}) and
(\ref{scalscalar}) that leads to the
$2+1$ dimensional $U(M-N)$ CS theory of
level $k$ that describes $M-N$
fractional M2-branes localized at the
core of $\mathbf{C^4/Z_k}$ and to
  $SU(N)$ L-BLG theory that describes
  $N$ M2-branes infinity far from
  singularity. Then it is natural
  that  these two theories are
  completely decoupled.
\vskip 4mm
 Note added: After submitting
the first version of this paper to
arXiv archive we were noticed by S.J.
Rey about his forthcoming paper
\cite{Rey} that has some overlap with
us.

\vskip .2in \noindent {\bf
Acknowledgements:} This work was
 supported by the Czech
Ministry of Education under Contract
No. MSM 0021622409.

\newpage

\end{document}